\documentclass[11pt]{article}
\usepackage{epstopdf}
\usepackage{color}
\usepackage{subfigure}
\usepackage{amsmath}
\usepackage{amssymb}
\usepackage{graphicx,color}
\usepackage{cite}
\usepackage{enumerate}
\usepackage{amsthm}
\usepackage{amsfonts,mathrsfs}
\usepackage{geometry}
\usepackage{ifthen}
\usepackage{graphicx}
\usepackage{float}
\usepackage{bm}
\usepackage[caption=false]{subfig}
\usepackage[utf8]{inputenc}

\begin{document}

\title{\bf Tighter sum uncertainty relations via $(\alpha,\beta,\gamma)$ weighted Wigner-Yanase-Dyson skew information }

\vskip0.1in
\author{\small Cong Xu$^1$, Zhaoqi Wu$^2$\thanks{Corresponding author. E-mail: wuzhaoqi\_conquer@163.com},
Shao-Ming Fei$^{1,3}$\thanks{Corresponding author. E-mail: feishm@cnu,edu.cn}\\
{\small\it  1. School of Mathematical Sciences, Capital Normal University, Beijing 100048, P R China}\\
{\small\it  2. Department of Mathematics, Nanchang University,
Nanchang 330031, P R China}\\
{\small\it  3. Max-Planck-Institute for Mathematics in the Sciences,
04103 Leipzig, Germany}}

\date{}
\maketitle

\noindent {\bf Abstract} {\small }\\
We establish tighter uncertainty relations for arbitrary finite observables via {\it $(\alpha,\beta,\gamma)$ weighted Wigner-Yanase-Dyson} ($(\alpha,\beta,\gamma)$WWYD) skew information. The results are also applicable to the {\it $(\alpha,\gamma)$ weighted Wigner-Yanase-Dyson} ($(\alpha,\gamma)$WWYD) skew information and the {\it weighted Wigner-Yanase-Dyson} (WWYD) skew information. We also present tighter lower bounds of quantum channels and unitary channels via {\it $(\alpha,\beta,\gamma)$ modified weighted Wigner-Yanase-Dyson} ($(\alpha,\beta,\gamma)$MWWYD) skew information. Detailed examples are provided to illustrate tightness of our uncertainty relations.

\vskip0.2in

\noindent {\bf 1. Introduction}\\
As one of the cornerstones of quantum mechanics, the uncertainty
principle has been widespread concerned since Heisenberg \cite{HW}
originally proposed the uncertainty principle in $1927$, which indicates that the position and momentum of a particle can not be precisely determined simultaneously. For arbitrary two observables $A$ and $B$ and a quantum state $|\psi\rangle$, the well-known Heisenberg-Robertson uncertainty relation \cite{RH} says that,
\begin{equation}\label{eq1}
\Delta A\Delta B\geq \frac{1}{2}|\langle\psi|[A,B]|\psi\rangle|,
\end{equation}
where $[A,B]=AB-BA$ and $\Delta
M=\sqrt{\langle\psi|{M}^2|\psi\rangle-{\langle\psi|M|\psi\rangle}^2}$
is the standard deviation of an observable $M$.

There are a host of methods to characterize uncertainty relations, such as entropy
\cite{DD,MHU,WSWA,WSYS,RAE}, variance \cite{GUDDER,DD1,DD2,SL}, successive measurements \cite{SMD} and majorization techniques \cite{PZRL,RLPZ,RL,FSGV}. In \cite{LUO5} Luo used the skew information to describe the uncertainty relation. The skew information of an observable $A$ with respect to a quantum state $\rho$ is given by \cite{WY},
\begin{align}\label{eq2}
\mathrm{I}_{\rho}(A)=-\frac{1}{2}\mathrm{Tr}
\left(\left[\sqrt{\rho},A\right]^2\right)
=\frac{1}{2}\left\|\left[\sqrt{\rho},A\right]\right\|^{2},
\end{align}
named as {\it Wigner-Yanase} (WY) skew information. Then a more general quantity was suggested by Dyson, called {\it Wigner-Yanase-Dyson} (WYD) skew information \cite{WY}. This quantity was further generalized in \cite{CL}, which is now termed as {\it generalized Wigner-Yanase-Dyson} (GWYD) skew information. Recently, the coherence and uncertainty relation via WY skew information, WYD skew information and GWYD skew information have been studied extensively \cite{LUO3,YANA1,YANA2,LUO9,XWF3,WU1,WU2,WU3,WU4,WU5,HWF1}.

Different from the WYD skew information, Furuichi et al. \cite{FURU1} generalized the WY skew information to {\it weighted Wigner-Yanase-Dyson} (WWYD) skew
information \cite{WU2} by considering the arithmetic mean of $\rho^\alpha$ and $\rho^{1-\alpha}$,
\begin{align}\label{eq3}
\mathrm{K}_{\rho}^{\alpha}(A)=-\frac{1}{2}\mathrm{Tr}
\left(\left[\frac{\rho^\alpha+\rho^{1-\alpha}}{2},A\right]^2\right)
=\frac{1}{2}\left\|\left[\frac{\rho^\alpha+\rho^{1-\alpha}}{2},
A\right]\right\|^{2},\,\,0\leq \alpha \leq 1.
\end{align}
A generalization of (\ref{eq3}) is given in \cite{CZL} for arbitrary operator $E$ (not necessarily Hermitian),
\begin{align}\label{eq4}
\mathrm{K}_{\rho}^{\alpha}(E)=-\frac{1}{2}\mathrm{Tr}\left(\left[\frac{\rho^\alpha
+\rho^{1-\alpha}}{2},E^{\dag}\right]\left[\frac{\rho^\alpha+\rho^{1-\alpha}}{2},E\right]\right)
=\frac{1}{2}\left\|\left[\frac{\rho^\alpha+\rho^{1-\alpha}}{2},
E\right]\right\|^{2} ,\,\,0\leq \alpha \leq 1,
\end{align}
which is termed as the {\it modified weighted Wigner-Yanase-Dyson}
(MWWYD) skew information in \cite{WU2}. Note that Eq. (\ref{eq4}) reduces to the Eq. (\ref{eq18}) in \cite {DD1} and Eq. (\ref{eq3}) reduces to the Eq. (\ref{eq2}) when $\alpha=\frac{1}{2}$, respectively.

Zhang \cite{Zhang} defined the {\it two-parameter extension of the Wigner-Yanase skew information} by replacing the arithmetic mean of $\rho^\alpha$ and
$\rho^{1-\alpha}$ with their convex combination,
\begin{align}\label{eq5}
\mathrm{K}_{\rho,\gamma}^{\alpha}(A)\notag
=&-\frac{1}{2}\mathrm{Tr}\left([(1-\gamma)\rho^\alpha+\gamma\rho^{1-\alpha},A]^{2}\right)\\
=&\frac{1}{2}\left\|\left[(1-\gamma)\rho^\alpha+\gamma\rho^{1-\alpha},
A\right]\right\|^{2} ~,~\,\,0\leq \alpha \leq 1~,\,\,0\leq \gamma
\leq 1,
\end{align}
which is called the {\it $(\alpha,\gamma)$ weighted
Wigner-Yanase-Dyson} ($(\alpha,\gamma)$ WWYD) skew information in
\cite{XWF1}. By replacing observables with arbitrary operators $E$, the authors in \cite{XWF2} obtained the {\it $(\alpha,\gamma)$ modified weighted Wigner-Yanase-Dyson} ($(\alpha,\gamma)$ MWWYD) skew information,
\begin{align}\label{eq6}
\mathrm{K}_{\rho,\gamma}^{\alpha}(E)\notag
=&-\frac{1}{2}\mathrm{Tr}([(1-\gamma)\rho^{\alpha}+\gamma\rho^{1-\alpha},E^{\dag}]
[(1-\gamma)\rho^{\alpha}+\gamma\rho^{1-\alpha},E])\\
=&\frac{1}{2}\left\|\left[(1-\gamma)\rho^\alpha+\gamma\rho^{1-\alpha},
E\right]\right\|^{2},~~\,\,0\leq \alpha \leq 1~,\,\,0\leq \gamma \leq 1.
\end{align}
Eq. (\ref{eq6}) reduces to Eq. (\ref{eq4}) and Eq. (\ref{eq5}) to Eq. (\ref{eq3}) when $\gamma=\frac{1}{2}$, respectively.

Motivated by the two-parameter extension Eq. (\ref{eq5}) and the Eq. (\ref{eq3}) in \cite{WU1}, we introduced the {\it $(\alpha,\beta,\gamma)$ weighted Wigner-Yanase-Dyson} ($(\alpha,\beta,\gamma)$ WWYD) skew information \cite{XWF2},
\begin{align}\label{eq7}
\mathrm{K}_{\rho,\gamma}^{\alpha,\beta}(A)\notag
=&-\frac{1}{2}\mathrm{Tr}([(1-\gamma)\rho^{\alpha}+\gamma\rho^{\beta},A]^{2}
\rho^{1-\alpha-\beta})\\
=&\frac{1}{2}\left\|\left[(1-\gamma)\rho^\alpha+\gamma\rho^\beta,
A\right]\rho^\frac{1-\alpha-\beta}{2}\right\|^{2},~~\alpha,\beta\geq
0,~\alpha+\beta\leq 1,0\leq \gamma \leq 1,
\end{align}
and its non-Hermitian extension the {\it $(\alpha,\beta,\gamma)$ modified
weighted Wigner-Yanase-Dyson} ($(\alpha,\beta,\gamma)$ MWWYD) skew information \cite{XWF2},
\begin{align}\label{eq8}
\mathrm{K}_{\rho,\gamma}^{\alpha,\beta}(E)\notag
=&-\frac{1}{2}\mathrm{Tr}([(1-\gamma)\rho^{\alpha}+\gamma\rho^{\beta},E^{\dag}]
[(1-\gamma)\rho^{\alpha}+\gamma\rho^{\beta},E]\rho^{1-\alpha-\beta})\\
=&\frac{1}{2}\left\|\left[(1-\gamma)\rho^\alpha+\gamma\rho^\beta,
E\right]\rho^\frac{1-\alpha-\beta}{2}\right\|^{2},~~~\alpha,\beta\geq
0,~\alpha+\beta\leq 1,0\leq\gamma \leq 1.
\end{align}
Here Eq. (\ref{eq8}) reduces to Eq. (\ref{eq6}) and Eq. (\ref{eq7}) reduces to Eq. (\ref{eq5}) when $\beta=1-\alpha$, respectively.

Quantum channels characterize the general evolutions of quantum systems \cite{BG,NC}. In particular, the unitary channels have potential applications in quantum information processing \cite{NC}. The sum uncertainty relations based on WY skew information have attracted considerable attention, and a plenty of results have been obtained \cite{CB3,FSS,ZL,ZWF1}. Recently, generalized uncertainty inequalities associated with metric-adjusted skew information for arbitrary finite quantum observables and quantum channels have been derived \cite{CAL,RRNL,ZWF2,HLTG}.

The paper is structured as follows. In Section 2, by using the operator norm inequalities, new uncertainty relations of observables are given in terms of the $(\alpha,\beta,\gamma)$ WWYD skew information. We present two distinct types of uncertainty relations of quantum channels with respect to the $(\alpha,\beta,\gamma)$ MWWYD skew information and derive an optimal lower bound in Section 3. The tighter uncertainty relation of unitary channels are presented in Section 4. We conclude with a summary in Section 5.

\medskip
\noindent {\bf 2. Sum uncertainty relations for arbitrary finite $N$ observables}

For arbitrary finite $N$ observables $A_1, A_2,\cdots,A_N$, Xu et al. \cite{XWF1} provided the following sum uncertainty relations,
\begin{align}\label{eq9}
\sum_{i=1}^{N} \mathrm{K}_{\rho,\gamma}^{\alpha,\beta}(A_i)
\geq&\frac{1}{N-2}\!\left[\sum_{1\leq i<j\leq N}\! \mathrm{K}_{\rho,\gamma}^{\alpha,\beta}(A_i+A_j)-\frac{1}{(N-1)^2}\!
 \left(\sum_{1\leq i<j\leq N}\!\sqrt{\mathrm{K}_{\rho,\gamma}^{\alpha,\beta}(A_i+A_j)}\right)^2\right]\!,
\end{align}
\begin{eqnarray}\label{eq10}
\sum_{i=1}^{N}{\mathrm{K}_{\rho,\gamma}^{\alpha,\beta}(A_i)}\geq \frac{1}{N}\mathrm{K}_{\rho,\gamma}^{\alpha,\beta}\left(\sum_{i=1}^ {N}A_i\right)+\frac{2}{N^2(N-1)}
\left(\sum_{1\leq i <j\leq N} \sqrt{\mathrm{K}_{\rho,\gamma}^{\alpha,\beta}(A_i-A_j)}\right)^2~,
\end{eqnarray}
and
\begin{align}\label{eq11}
\sum_{i=1}^{N}{\mathrm{K}_{\rho,\gamma}^{\alpha,\beta}(A_i)}
\geq& \mathop{\mathrm{max}}\limits_{x\in\{0,1\}}\frac{1}{2(N-1)}
\left[\frac{2}{N(N-1)}\left(\sum_{1\leq i<j\leq N} \sqrt{\mathrm{K}_{\rho,\gamma}^{\alpha,\beta}(A_i+(-1)^{x}A_j)}\right)^2 \right.
\nonumber\\
&\left.+\sum_{1\leq i<j\leq N} \mathrm{K}_{\rho,\gamma}^{\alpha,\beta}(A_i+(-1)^{x+1}A_j)\right],
\end{align}
where $\alpha,\beta\geq 0,~\alpha+\beta\leq 1,~0\leq \gamma \leq 1$, $N>2$ for the inequality (\ref{eq9}) and $N\geq2$ for the inequalities (\ref{eq10}-\ref{eq11}).
For convenience we denote by $Lb_1$, $Lb_2$ and $Lb_3$ the right hand sides of (\ref{eq9}), (\ref{eq10}) and (\ref{eq11}), respectively.

The following relations are given in the Appendix B in \cite{HLTG},
\begin{align}\label{eq16}
\sum_{i=1}^{N} \| u_i\|^2
&\geq\frac{1}{MN+(N-2)L}\left\{\frac{2L}{N(N-1)}\left(\sum_{1\leq i<j\leq N}\| u_i+u_j\|\right)^2+\right.
\nonumber\\
&\left.M\sum_{1\leq i<j\leq N}\| u_i-u_j\|^2+
(M-L)\left\|\sum_{i=1}^{N}u_i\right\|^2\right\},
\end{align}
for $M\geq L>0$,
\begin{align}\label{eq17}
\sum_{i=1}^{N} \| u_i\|^2
&\geq\frac{1}{MN+(N-2)L}\left\{\frac{2M}{N(N-1)}\left(\sum_{1\leq i<j\leq N}\| u_i-u_j\|\right)^2+\right.
\nonumber\\
&\left.L\sum_{1\leq i<j\leq N}\| u_i+u_j\|^2+
(M-L)\left\|\sum_{i=1}^{N}u_i\right\|^2\right\},
\end{align}
for $L\geq M>0$,
\begin{align}\label{eq18}
\sum_{i=1}^{N} \| u_i\|^2
&\geq\frac{1}{MN+(N-2)L}\left\{\frac{M-L}{(N-1)^2}\left(\sum_{1\leq i<j\leq N}\| u_i+u_j\|\right)^2+\right.
\nonumber\\
&\left.M\sum_{1\leq i<j\leq N}\| u_i-u_j\|^2+L\sum_{1\leq i<j\leq N}\| u_i+u_j\|^2
\right\}
\end{align}
for arbitrary $L>M>0$. By replacing $u_i$ and $u_j$ with $\left[(1-\gamma)\rho^\alpha+\gamma\rho^\beta,A_i\right]\rho^\frac{1-\alpha-\beta}{2}$ and
$\left[(1-\gamma)\rho^\alpha+\gamma\rho^\beta,A_j\right]\rho^\frac{1-\alpha-\beta}{2}$,
respectively, in the above inequalities, we obtain the following inequalities.

{\bf Theorem 1} For arbitrary finite $N$ observables $A_1, A_2,\cdots,A_N$ ($N\geq2$), we have the following sum uncertainty relation via $(\alpha,\beta,\gamma)$ WWYD skew information,
\begin{align}\label{eq12}
\sum_{i=1}^{N}\mathrm{K}_{\rho,\gamma}^{\alpha,\beta}(A_i)\geq \mathop{\mathrm{max}}\{\overline{Lb}_1,\overline{Lb}_2,\overline{Lb}_3\},
\end{align}
where
\begin{align}\label{eq13}
\overline{Lb}_1
&=\frac{1}{MN+(N-2)L}\left\{\frac{2L}{N(N-1)}\left(\sum_{1\leq i<j\leq N}\sqrt{\mathrm{K}_{\rho,\gamma}^{\alpha,\beta}(A_i+A_j)}\right)^2+\right.
\nonumber\\
&\left.M\sum_{1\leq i<j\leq N}\mathrm{K}_{\rho,\gamma}^{\alpha,\beta}(A_i-A_j)+
(M-L)\mathrm{K}_{\rho,\gamma}^{\alpha,\beta}
\left(\sum_{i=1}^{N}A_{i}\right)\right\},
\end{align}
\begin{align}\label{eq14}
\overline{Lb}_2
&=\frac{1}{MN+(N-2)L}\left\{\frac{2M}{N(N-1)}\left(\sum_{1\leq i<j\leq N}\sqrt{\mathrm{K}_{\rho,\gamma}^{\alpha,\beta}(A_i-A_j)}\right)^2+\right.
\nonumber\\
&\left.L\sum_{1\leq i<j\leq N}\mathrm{K}_{\rho,\gamma}^{\alpha,\beta}(A_i+A_j)+
(M-L)\mathrm{K}_{\rho,\gamma}^{\alpha,\beta}
\left(\sum_{i=1}^{N}A_{i}\right)\right\},
\end{align}
\begin{align}\label{eq15}
\overline{Lb}_3
&=\frac{1}{MN+(N-2)L}\left\{\frac{M-L}{(N-1)^2}\left(\sum_{1\leq i<j\leq N}\sqrt{\mathrm{K}_{\rho,\gamma}^{\alpha,\beta}(A_i+A_j)}\right)^2 \right.
\nonumber\\
&\left.+
M\sum_{1\leq i<j\leq N}\mathrm{K}_{\rho,\gamma}^{\alpha,\beta}(A_i-A_j)+L\sum_{1\leq i<j\leq N}\mathrm{K}_{\rho,\gamma}^{\alpha,\beta}(A_i+A_j)
\right\},
\end{align}
$\alpha,\beta\geq 0$, $\alpha+\beta\leq 1$, $0\leq \gamma \leq 1$, the parameters $L$, $M$ in the expressions of $\overline{Lb}_1$, $\overline{Lb}_2$ and $\overline{Lb}_3$ satisfy $M\geq L>0$, $L\geq M>0$ and $L>M>0$, respectively.

For convenience, we denote $\overline{Lb}=\mathop{\mathrm{max}}\{\overline{Lb}_1,\overline{Lb}_2,\overline{Lb}_3\}$
the right hand side of (\ref{eq12}).
In \cite{HLTG} Li et al. proved that (\ref{eq16}), (\ref{eq17}) and (\ref{eq18}) are strictly tighter than those of norm inequalities related to (\ref{eq9}), (\ref{eq10}) and (\ref{eq11}) for appropriate $M$ and $L$. If we take $M=L$, $\overline{Lb}_1$ and $\overline{Lb}_2$ reduces to the cases of $Lb_3$. For fixed $N\,(\geq2)$, larger $M$ and smaller $L$ give rise to larger right hand sides of the inequalities (\ref{eq16}) and (\ref{eq18}). Conversely, smaller $M$ and larger $L$ result in larger right hand side of the inequality (\ref{eq17}).

{\bf Corollary 1} For finite $N$ observables $A_1, A_2, \cdots, A_N$ ($N\geq2$), the sum uncertainty relations with respect to WWYD skew information are given by
\begin{align}\label{eq19}
\sum_{i=1}^{N}{\mathrm{K}_{\rho}^{\alpha}(A_i)}
&\geq\frac{1}{MN+(N-2)L}\left\{\frac{2L}{N(N-1)}\left(\sum_{1\leq i<j\leq N}\sqrt{\mathrm{K}_{\rho}^{\alpha}(A_i+A_j)}\right)^2+\right.
\nonumber\\
&\left.M\sum_{1\leq i<j\leq N}\mathrm{K}_{\rho}^{\alpha}(A_i-A_j)+
(M-L)\mathrm{K}_{\rho}^{\alpha}
\left(\sum_{i=1}^{N}A_{i}\right)\right\},
\end{align}
\begin{align}\label{eq20}
\sum_{i=1}^{N}{\mathrm{K}_{\rho}^{\alpha}(A_i)}
&\geq\frac{1}{MN+(N-2)L}\left\{\frac{2M}{N(N-1)}\left(\sum_{1\leq i<j\leq N}\sqrt{\mathrm{K}_{\rho}^{\alpha}(A_i-A_j)}\right)^2+\right.
\nonumber\\
&\left.L\sum_{1\leq i<j\leq N}\mathrm{K}_{\rho}^{\alpha}(A_i+A_j)+
(M-L)\mathrm{K}_{\rho}^{\alpha}
\left(\sum_{i=1}^{N}A_{i}\right)\right\}
\end{align}
and
\begin{align}\label{eq21}
\sum_{i=1}^{N}{\mathrm{K}_{\rho}^{\alpha}(A_i)}
&\geq\frac{1}{MN+(N-2)L}\left\{\frac{M-L}{(N-1)^2}\left(\sum_{1\leq i<j\leq N}\sqrt{\mathrm{K}_{\rho}^{\alpha}(A_i+A_j)}\right)^2 \right.
\nonumber\\
&\left.+
M\sum_{1\leq i<j\leq N}\mathrm{K}_{\rho}^{\alpha}(A_i-A_j)+L\sum_{1\leq i<j\leq N}\mathrm{K}_{\rho}^{\alpha}(A_i+A_j)
\right\},
\end{align}
where $0\leq\alpha\leq 1$, the parameters $L$, $M$ in (\ref{eq19}), (\ref{eq20}) and (\ref{eq21}) satisfy $M\geq L>0$, $L\geq M>0$ and $L>M>0$, respectively.

Denote $rhs19$, $rhs20$ and $rhs21$ the right hand sides of the inequalities (\ref{eq19}), (\ref{eq20}) and (\ref{eq21}), respectively. Corollary 1 implies that
$\sum_{i=1}^{N}{\mathrm{K}_{\rho}^{\alpha}(A_i)}\geq \max \{rhs19, rhs20,rhs21\}$. In particular, (\ref{eq33}), (\ref{eq34}) and (\ref{eq35}) in \cite{HLTG} are just the special cases of the inequalities (\ref{eq19}), (\ref{eq20}) and (\ref{eq21}) for $\alpha=\frac{1}{2}$, respectively. Next we prove that our new lower bound $\overline{Lb}$ is tighter than the existing ones by a detailed example. We consider the WWYD skew information as a special case, and take $M=2$, $L=1$ for $\overline{Lb}_1$, and $M=1$, $L=2$ for $\overline{Lb}_2$ and $\overline{Lb}_3$.

{\bf Example 1} Consider the pure state
$\rho=\frac{1}{2}(\mathbf{1}+\mathbf{r}\cdot\bm{\sigma})$, where $\mathbf{1}$ is the $2\times2$ identity matrix, $\mathbf{r}=(\frac{\sqrt{2}}{2}\cos\theta,\frac{\sqrt{2}}{2}\sin\theta,\frac{\sqrt{2}}{2})$ is the Bloch vector satisfying
$|\mathbf{r}|\leq 1$, $\bm{\sigma}=(\sigma_1,\sigma_2,\sigma_3)$
with $\sigma_j$ $(j=1,2,3)$ the Pauli matrices, and
$\mathbf{r}\cdot\bm{\sigma}=\sum^3_{j=1}r_j\sigma_j$. We compare $\overline{Lb}$ with $Lb_1$, $Lb_2$ and $Lb_3$ for any $\alpha$ and $\alpha=\frac{1}{3}$, respectively.
It can be seen that $\overline{Lb}$ is tighter than $Lb_1$, $Lb_2$ and $Lb_3$ for arbitrary $\alpha$, see Figure $1$.
\begin{figure}[H]\centering
\subfigure[]
{\begin{minipage}[Xu-Cong-uncertainty1a]{0.49\linewidth}
\includegraphics[width=0.98\textwidth]{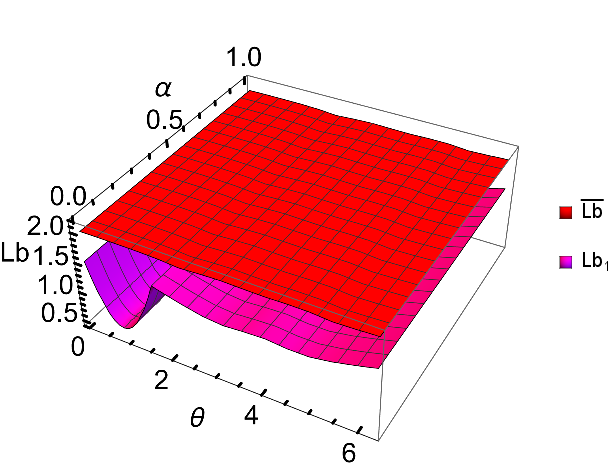}
\end{minipage}}
\subfigure[]
{\begin{minipage}[Xu-Cong-uncertainty1b]{0.49\linewidth}
\includegraphics[width=0.98\textwidth]{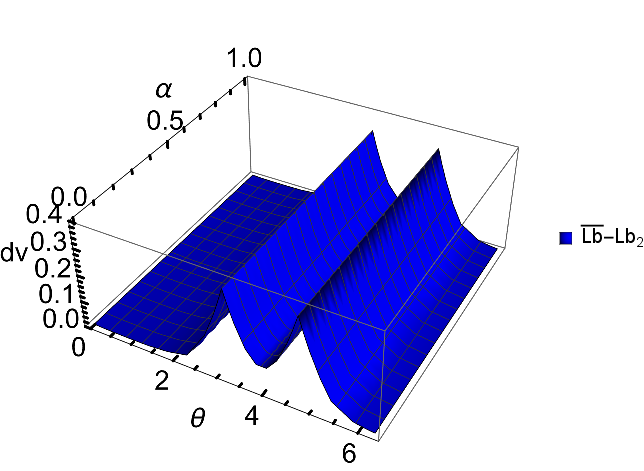}
\end{minipage}}
\subfigure[]
{\begin{minipage}[Xu-Cong-uncertainty1c]{0.49\linewidth}
\includegraphics[width=0.98\textwidth]{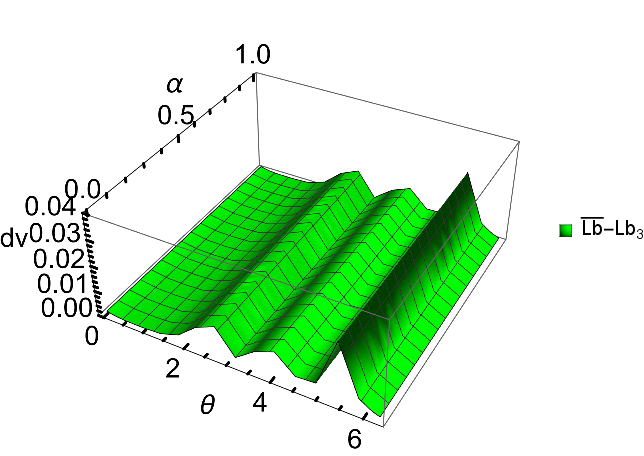}
\end{minipage}}
\subfigure[]
{\begin{minipage}[Xu-Cong-uncertainty1d]{0.49\linewidth}
\includegraphics[width=0.98\textwidth]{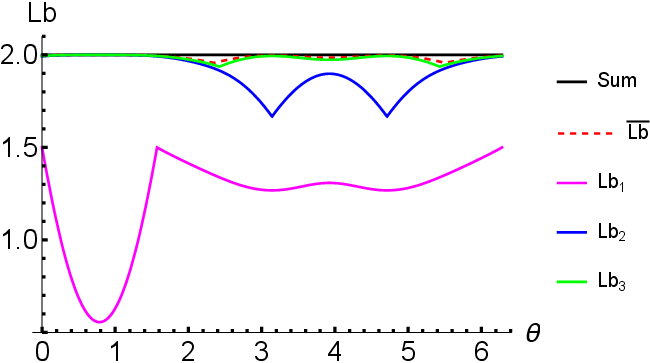}
\end{minipage}}
\caption{{Comparison of the lower bound (Lb) $\overline{Lb}$ with $Lb_1$, $Lb_2$ and $Lb_3$. (a) The red surface, the magenta surface represent $\overline{Lb}$ and $Lb_1$, respectively. (b) The blue surface represents the difference value (dv) between $\overline{Lb}$ and $Lb_2$. (c) The green surface represents the difference value between $\overline{Lb}$ and $Lb_3$. (d) For fixed $\alpha=\frac{1}{3}$, the black, red dashed, magenta, blue and green curves represent the summation $\mathrm{K}_{\rho}^{\frac{1}{3}}(\sigma_1)+\mathrm{K}_{\rho}^{\frac{1}{3}}(\sigma_2)
+\mathrm{K}_{\rho}^{\frac{1}{3}}(\sigma_3)$, $\overline{Lb}$, $Lb_1$, $Lb_2$ and $Lb_3$, respectively.
\label{fig:Fig1}}}
\end{figure}

\medskip
\noindent {\bf 3. Sum uncertainty relations for finite quantum channels}
\\
In this section, we give two different types of uncertainty relations associated with arbitrary finite number of quantum channels based on $(\alpha,\beta,\gamma)$ MWWYD skew information. We derive an optimal lower bound and show that our bounds are tighter than the existing ones by a detailed example.

Let $\Phi$ be a quantum channel with Kraus representation, $\Phi(\rho)=\sum_{i=1}^{n}E_i\rho E_i^{\dag}$. In \cite{XWF1} the authors have presented the uncertainty quantification with respect to a channel $\Phi$ via $(\alpha,\beta,\gamma)$ MWWYD skew information,
\begin{equation}\label{eq22}
\mathrm{K}_{\rho,\gamma}^{\alpha,\beta}(\Phi)=\sum_{i=1}^{n}
\mathrm{K}_{\rho,\gamma}^{\alpha,\beta}(E_i),
\end{equation}
where $\alpha,\beta\geq0,~\alpha+\beta\leq 1,0\leq \gamma \leq 1$.
For arbitrary $N$ quantum channels ,$\Phi_{1},\cdots,\Phi_N$ with Kraus representations $\Phi_t(\rho)=\sum_{i=1}^{n}E_{i}^{t}\rho (E_{i}^{t})^\dag, ~t=1,2,\cdots,N$ ($N>2$ for the inequality (\ref{eq23}) and $N\geq2$ for the inequalities (\ref{eq24}-\ref{eq25})). Xu et al.\cite{XWF1} gave the following sum uncertainty quantifications associated with channels,
\begin{align}\label{eq23}
\sum_{t=1}^{N}\mathrm{K}_{\rho,\gamma}^{\alpha,\beta}(\Phi_t)
\geq& \mathop{\mathrm{max}}\limits_{\pi_t,\pi_s\in S_n}\frac{1}{N-2}\left\{\sum_{1\leq t<s\leq N}\sum_{i=1}^{n}\mathrm{K}_{\rho,\gamma}^{\alpha,\beta}(E_{\pi_{t}(i)}^{t}+E_{\pi_{s}(i)}^{s}) \right.
\nonumber\\
&\left.-\frac{1}{(N-1)^{2}}\left[\sum_{i=1}^{n}\left(\sum_{1\leq t<s\leq N}\sqrt{\mathrm{K}_{\rho,\gamma}^{\alpha,\beta}(E_{\pi_{t}(i)}^{t}
+E_{\pi_{s}(i)}^{s})}\right)^{2}\right]\right\},
\end{align}
\begin{align}\label{eq24}
\sum_{t=1}^{N}\mathrm{K}_{\rho,\gamma}^{\alpha,\beta}(\Phi_t)
\geq& \mathop{\mathrm{max}}\limits_{\pi_t,\pi_s\in S_n}\left\{\frac{1}{N}\sum_{i=1}^{n}\mathrm{K}_{\rho,\gamma}^{\alpha,\beta}
\left(\sum_{t=1}^{N}E_{\pi_{t}(i)}^{t}\right) \right.
\nonumber\\
&\left.+\frac{2}{N^{2}(N-1)}\left[\sum_{i=1}^{n}\left(\sum_{1\leq t<s\leq N}\sqrt{\mathrm{K}_{\rho,\gamma}^{\alpha,\beta}(E_{\pi_{t}(i)}^{t}
-E_{\pi_{s}(i)}^{s})}\right)^{2}\right]\right\},
\end{align}
\begin{align}\label{eq25}
\sum_{t=1}^{N}\mathrm{K}_{\rho,\gamma}^{\alpha,\beta}(\Phi_t)
\geq& \mathop{\mathrm{max}}\limits_{\pi_t,\pi_s\in S_n}\frac{1}{2(N-1)}\left\{\frac{2}{N(N-1)}\left[\sum_{i=1}^{n}\left(\sum_{1\leq t<s\leq N}\sqrt{\mathrm{K}_{\rho,\gamma}^{\alpha,\beta}(E_{\pi_{t}(i)}^{t}
\pm E_{\pi_{s}(i)}^{s})}\right)^2\right] \right.
\nonumber\\
&\left.+\sum_{1\leq t<s\leq N}\sum_{i=1}^{n}\mathrm{K}_{\rho,\gamma}^{\alpha,\beta}(E_{\pi_{t}(i)}^{t}
\mp E_{\pi_{s}(i)}^{s})\right\},
\end{align}
where $\alpha,\beta\geq
0,~\alpha+\beta\leq 1,~0\leq \gamma \leq 1$, $S_n$ is the $n$-element permutation group and $\pi_{t},\pi_{s}\in S_n$ are arbitrary $n$-element permutations.
We denote by $LB_1$, $LB_2$ and $LB_3$ the right hand sides of (\ref{eq23}), (\ref{eq24}) and (\ref{eq25}), respectively.

{\bf Theorem 2} Let $\Phi_{1},\cdots,\Phi_N$ be $N$ quantum
channels with Kraus representations
$\Phi_t(\rho)=\sum_{i=1}^{n}E_{i}^{t}\rho (E_{i}^{t})^\dag, ~t=1,2,\cdots,N$ ($N\geq2$). We have
\begin{align}\label{eq26}
\sum_{t=1}^{N}\mathrm{K}_{\rho,\gamma}^{\alpha,\beta}(\Phi_t)\geq \mathop{\mathrm{max}}\{\overline{LB}_1,\overline{LB}_2,\overline{LB}_3\},
\end{align}
where
\begin{align}\label{eq27}
\overline{LB}_1
&=\mathop{\mathrm{max}}\limits_{\pi_t,\pi_s\in S_n}\frac{1}{MN+(N-2)L}\left\{\frac{2L}{N(N-1)}\left[\sum_{i=1}^{n}\left(\sum_{1\leq t<s\leq N}\sqrt{\mathrm{K}_{\rho,\gamma}^{\alpha,\beta}(E_{\pi_{t}(i)}^{t}
+E_{\pi_{s}(i)}^{s})}\right)^{2}\right]\right.
\nonumber\\
&\left.+M\sum_{1\leq t<s\leq N}\sum_{i=1}^{n}\mathrm{K}_{\rho,\gamma}^{\alpha,\beta}(E_{\pi_{t}(i)}^{t}-E_{\pi_{s}(i)}^{s}) +
(M-L)\sum_{i=1}^{n}\mathrm{K}_{\rho,\gamma}^{\alpha,\beta}
\left(\sum_{t=1}^{N}E_{\pi_{t}(i)}^{t}\right)\right\},
\end{align}
\begin{align}\label{eq28}
\overline{LB}_2
&=\mathop{\mathrm{max}}\limits_{\pi_t,\pi_s\in S_n}\frac{1}{MN+(N-2)L}\left\{\frac{2M}{N(N-1)}\left[\sum_{i=1}^{n}\left(\sum_{1\leq t<s\leq N}\sqrt{\mathrm{K}_{\rho,\gamma}^{\alpha,\beta}(E_{\pi_{t}(i)}^{t}
-E_{\pi_{s}(i)}^{s})}\right)^{2}\right]\right.
\nonumber\\
&\left.+L\sum_{1\leq t<s\leq N}\sum_{i=1}^{n}\mathrm{K}_{\rho,\gamma}^{\alpha,\beta}(E_{\pi_{t}(i)}^{t}+E_{\pi_{s}(i)}^{s}) +
(M-L)\sum_{i=1}^{n}\mathrm{K}_{\rho,\gamma}^{\alpha,\beta}
\left(\sum_{t=1}^{N}E_{\pi_{t}(i)}^{t}\right)\right\},
\end{align}
\begin{align}\label{eq29}
\overline{LB}_3
&=\mathop{\mathrm{max}}\limits_{\pi_t,\pi_s\in S_n}\frac{1}{MN+(N-2)L}\left\{\frac{M-L}{(N-1)^2}\left[\sum_{i=1}^{n}\left(\sum_{1\leq t<s\leq N}\sqrt{\mathrm{K}_{\rho,\gamma}^{\alpha,\beta}(E_{\pi_{t}(i)}^{t}
+E_{\pi_{s}(i)}^{s})}\right)^{2}\right]\right.
\nonumber\\
&\left.+L\sum_{1\leq t<s\leq N}\sum_{i=1}^{n}\mathrm{K}_{\rho,\gamma}^{\alpha,\beta}(E_{\pi_{t}(i)}^{t}+E_{\pi_{s}(i)}^{s}) +
M\sum_{1\leq t<s\leq N}\sum_{i=1}^{n}\mathrm{K}_{\rho,\gamma}^{\alpha,\beta}(E_{\pi_{t}(i)}^{t}-E_{\pi_{s}(i)}^{s}) \right\},
\end{align}
$\alpha,\beta\geq 0$, $\alpha+\beta\leq 1$, $0\leq \gamma \leq 1$, $\pi_{t},\pi_{s}\in S_n$ are arbitrary $n$-element permutations, the parameters $L$, $M$ in $\overline{LB}_1$, $\overline{LB}_2$ and $\overline{LB}_3$ satisfy $M\geq L>0$, $L\geq M>0$ and $L>M>0$, respectively.

\noindent\textit{Proof} Setting $u_i=\left[(1-\gamma)\rho^\alpha+\gamma\rho^\beta,E_i\right]\rho^\frac{1-\alpha-\beta}{2}$ and $u_j=\left[(1-\gamma)\rho^\alpha+\gamma\rho^\beta,E_j\right]\rho^\frac{1-\alpha-\beta}{2}$,
in (\ref{eq16}), (\ref{eq17}) and (\ref{eq18}), from (\ref{eq8}) we have
\begin{align*}
\sum_{t=1}^{N}\mathrm{K}_{\rho,\gamma}^{\alpha,\beta}(E_{\pi_{t}(i)}^{t})
&\geq\mathop{\mathrm{max}}\limits_{\pi_t,\pi_s\in S_n}\frac{1}{MN+(N-2)L}\left\{\frac{2L}{N(N-1)}\left[\left(\sum_{1\leq t<s\leq N}\sqrt{\mathrm{K}_{\rho,\gamma}^{\alpha,\beta}(E_{\pi_{t}(i)}^{t}
+E_{\pi_{s}(i)}^{s})}\right)^{2}\right]\right.
\nonumber\\
&\left.+M\sum_{1\leq t<s\leq N}\mathrm{K}_{\rho,\gamma}^{\alpha,\beta}(E_{\pi_{t}(i)}^{t}-E_{\pi_{s}(i)}^{s}) +
(M-L)\mathrm{K}_{\rho,\gamma}^{\alpha,\beta}
\left(\sum_{t=1}^{N}E_{\pi_{t}(i)}^{t}\right)\right\},
\end{align*}
for $M\geq L>0$,
\begin{align*}
\sum_{t=1}^{N}\mathrm{K}_{\rho,\gamma}^{\alpha,\beta}(E_{\pi_{t}(i)}^{t})
&\geq\mathop{\mathrm{max}}\limits_{\pi_t,\pi_s\in S_n}\frac{1}{MN+(N-2)L}\left\{\frac{2M}{N(N-1)}\left[\left(\sum_{1\leq t<s\leq N}\sqrt{\mathrm{K}_{\rho,\gamma}^{\alpha,\beta}(E_{\pi_{t}(i)}^{t}
-E_{\pi_{s}(i)}^{s})}\right)^{2}\right]\right.
\nonumber\\
&\left.+L\sum_{1\leq t<s\leq N}\mathrm{K}_{\rho,\gamma}^{\alpha,\beta}(E_{\pi_{t}(i)}^{t}+E_{\pi_{s}(i)}^{s}) +
(M-L)\mathrm{K}_{\rho,\gamma}^{\alpha,\beta}
\left(\sum_{t=1}^{N}E_{\pi_{t}(i)}^{t}\right)\right\},
\end{align*}
for $L\geq M>0$,
\begin{align*}
\sum_{t=1}^{N}\mathrm{K}_{\rho,\gamma}^{\alpha,\beta}(E_{\pi_{t}(i)}^{t})
&\geq\mathop{\mathrm{max}}\limits_{\pi_t,\pi_s\in S_n}\frac{1}{MN+(N-2)L}\left\{\frac{M-L}{(N-1)^2}\left[\left(\sum_{1\leq t<s\leq N}\sqrt{\mathrm{K}_{\rho,\gamma}^{\alpha,\beta}(E_{\pi_{t}(i)}^{t}
+E_{\pi_{s}(i)}^{s})}\right)^{2}\right]\right.
\nonumber\\
&\left.+L\sum_{1\leq t<s\leq N}\mathrm{K}_{\rho,\gamma}^{\alpha,\beta}(E_{\pi_{t}(i)}^{t}+E_{\pi_{s}(i)}^{s}) +
M\sum_{1\leq t<s\leq N}\mathrm{K}_{\rho,\gamma}^{\alpha,\beta}(E_{\pi_{t}(i)}^{t}-E_{\pi_{s}(i)}^{s}) \right\},
\end{align*}
for $L>M>0$. Summing over the index $i$ we obtain Theorem 2. $\Box$

{\bf Corollary 2} Let $\Phi_{1},\cdots,\Phi_N$ be $N$ quantum channels with Kraus representations $\Phi_t(\rho)=\sum_{i=1}^{n}E_{i}^{t}\rho (E_{i}^{t})^\dag, ~t=1,2,\cdots,N$ ($N\geq2$), we have the sum uncertainty relations with respect to WWYD skew information,
\begin{align}\label{eq30}
\sum_{t=1}^{N}\mathrm{K}_{\rho}^{\alpha}(\Phi_t)
&\geq\mathop{\mathrm{max}}\limits_{\pi_t,\pi_s\in S_n}\frac{1}{MN+(N-2)L}\left\{\frac{2L}{N(N-1)}\left[\sum_{i=1}^{n}\left(\sum_{1\leq t<s\leq N}\sqrt{\mathrm{K}_{\rho}^{\alpha}(E_{\pi_{t}(i)}^{t}
+E_{\pi_{s}(i)}^{s})}\right)^{2}\right]\right.
\nonumber\\
&\left.+M\sum_{1\leq t<s\leq N}\sum_{i=1}^{n}\mathrm{K}_{\rho}^{\alpha}(E_{\pi_{t}(i)}^{t}-E_{\pi_{s}(i)}^{s}) +
(M-L)\sum_{i=1}^{n}\mathrm{K}_{\rho}^{\alpha}
\left(\sum_{t=1}^{N}E_{\pi_{t}(i)}^{t}\right)\right\},
\end{align}
\begin{align}\label{eq31}
\sum_{t=1}^{N}\mathrm{K}_{\rho}^{\alpha}(\Phi_t)
&\geq\mathop{\mathrm{max}}\limits_{\pi_t,\pi_s\in S_n}\frac{1}{MN+(N-2)L}\left\{\frac{2M}{N(N-1)}\left[\sum_{i=1}^{n}\left(\sum_{1\leq t<s\leq N}\sqrt{\mathrm{K}_{\rho}^{\alpha}(E_{\pi_{t}(i)}^{t}
-E_{\pi_{s}(i)}^{s})}\right)^{2}\right]\right.
\nonumber\\
&\left.+L\sum_{1\leq t<s\leq N}\sum_{i=1}^{n}\mathrm{K}_{\rho}^{\alpha}(E_{\pi_{t}(i)}^{t}+E_{\pi_{s}(i)}^{s}) +
(M-L)\sum_{i=1}^{n}\mathrm{K}_{\rho}^{\alpha}
\left(\sum_{t=1}^{N}E_{\pi_{t}(i)}^{t}\right)\right\},
\end{align}
\begin{align}\label{eq32}
\sum_{t=1}^{N}\mathrm{K}_{\rho}^{\alpha}(\Phi_t)
&\geq\mathop{\mathrm{max}}\limits_{\pi_t,\pi_s\in S_n}\frac{1}{MN+(N-2)L}\left\{\frac{M-L}{(N-1)^2}\left[\sum_{i=1}^{n}\left(\sum_{1\leq t<s\leq N}\sqrt{\mathrm{K}_{\rho}^{\alpha}(E_{\pi_{t}(i)}^{t}
+E_{\pi_{s}(i)}^{s})}\right)^{2}\right]\right.
\nonumber\\
&\left.+L\sum_{1\leq t<s\leq N}\sum_{i=1}^{n}\mathrm{K}_{\rho}^{\alpha}(E_{\pi_{t}(i)}^{t}+E_{\pi_{s}(i)}^{s}) +
M\sum_{1\leq t<s\leq N}\sum_{i=1}^{n}\mathrm{K}_{\rho}^{\alpha}(E_{\pi_{t}(i)}^{t}-E_{\pi_{s}(i)}^{s}) \right\},
\end{align}
where $0\leq\alpha\leq 1$, the parameters $L$, $M$ in $(\ref{eq30})$, $(\ref{eq31})$ and $(\ref{eq32})$ satisfy $M\geq L>0$, $L\geq M>0$ and $L>M>0$, respectively.

Thus we have $\sum_{i=1}^{N}{\mathrm{K}_{\rho}^{\alpha}(\Phi_t)}\geq \max \{rhs30, rhs31, rhs32\}$, where $rhs30$, $rhs31$ and $rhs32$ represent the right hand sides of inequalities (\ref{eq30}), (\ref{eq31}) and (\ref{eq32}), respectively.

Let $\Phi$ be a quantum channel with Kraus representation,
$\Phi(\rho)=\sum_{i=1}^{n}E_i\rho E_i^{\dag}$. The uncertainty relation of channel $\Phi$ via ($\alpha,\beta,\gamma$) MWWYD skew information can also be written as \cite{XWF2},
\begin{align}\label{eq33}
\mathrm{K}_{\rho,\gamma}^{\alpha,\beta}(\Phi)=\frac{1}{2}\mathrm{Tr}(u^\dag u)=\frac{1}{2}\|u\|^2,
\end{align}
where $\alpha,\beta\geq0$, $\alpha+\beta\leq 1$, $0\leq \gamma \leq 1$, $u=(\left[(1-\gamma)\rho^\alpha+\gamma\rho^\beta,
E_1\right]\rho^\frac{1-\alpha-\beta}{2},\left[(1-\gamma)\rho^\alpha+\gamma\rho^\beta \right.$\\ $\left.  ,E_2\right]\rho^\frac{1-\alpha-\beta}{2},\cdots,\left[(1-\gamma)\rho^\alpha+
\gamma\rho^\beta, E_n\right]\rho^\frac{1-\alpha-\beta}{2})$. Therefore, by using the inequalities (\ref{eq16}), (\ref{eq17}) and (\ref{eq18}) with $\|u_t\|^2=2\mathrm{K}_{\rho,\gamma}^{\alpha,\beta}(\Phi_t)$,
$\|u_t\pm u_s\|^2=2\sum_{i=1}^{n}
\mathrm{K}_{\rho,\gamma}^{\alpha,\beta}(E_{\pi_{t}(i)}^{t}\pm E_{\pi_{s}(i)}^{s})$, we have the following theorem.

{\bf Theorem 3} Let $\Phi_{1},\cdots,\Phi_N$ be $N$ quantum
channels with Kraus representations
$\Phi_t(\rho)=\sum_{i=1}^{n}E_{i}^{t}\rho (E_{i}^{t})^\dag, ~t=1,2,\cdots,N$ ($N\geq2$). We have
\begin{align}\label{eq34}
\sum_{t=1}^{N}\mathrm{K}_{\rho,\gamma}^{\alpha,\beta}(\Phi_t)\geq \mathop{\mathrm{max}}\{\overline{LB}1,\overline{LB}2,\overline{LB}3\},
\end{align}
where
\begin{align}\label{eq35}
\overline{LB}1
&=\mathop{\mathrm{max}}\limits_{\pi_t,\pi_s\in S_n}\frac{1}{MN+(N-2)L}\left\{\frac{2L}{N(N-1)}\left[\sum_{1\leq t<s\leq N}\sqrt{\sum_{i=1}^{n}\mathrm{K}_{\rho,\gamma}^{\alpha,\beta}(E_{\pi_{t}(i)}^{t}
+E_{\pi_{s}(i)}^{s})}\right]^{2}\right.
\nonumber\\
&\left.+M\sum_{1\leq t<s\leq N}\sum_{i=1}^{n}\mathrm{K}_{\rho,\gamma}^{\alpha,\beta}(E_{\pi_{t}(i)}^{t}-E_{\pi_{s}(i)}^{s}) +
(M-L)\sum_{i=1}^{n}\mathrm{K}_{\rho,\gamma}^{\alpha,\beta}
\left(\sum_{t=1}^{N}E_{\pi_{t}(i)}^{t}\right)\right\},
\end{align}
\begin{align}\label{eq36}
\overline{LB}2
&=\mathop{\mathrm{max}}\limits_{\pi_t,\pi_s\in S_n}\frac{1}{MN+(N-2)L}\left\{\frac{2M}{N(N-1)}\left[\sum_{1\leq t<s\leq N}\sqrt{\sum_{i=1}^{n}\mathrm{K}_{\rho,\gamma}^{\alpha,\beta}(E_{\pi_{t}(i)}^{t}
-E_{\pi_{s}(i)}^{s})}\right]^{2}\right.
\nonumber\\
&\left.+L\sum_{1\leq t<s\leq N}\sum_{i=1}^{n}\mathrm{K}_{\rho,\gamma}^{\alpha,\beta}(E_{\pi_{t}(i)}^{t}+E_{\pi_{s}(i)}^{s}) +
(M-L)\sum_{i=1}^{n}\mathrm{K}_{\rho,\gamma}^{\alpha,\beta}
\left(\sum_{t=1}^{N}E_{\pi_{t}(i)}^{t}\right)\right\},
\end{align}
\begin{align}\label{eq37}
\overline{LB}3
&=\mathop{\mathrm{max}}\limits_{\pi_t,\pi_s\in S_n}\frac{1}{MN+(N-2)L}\left\{\frac{M-L}{(N-1)^2}\left[\sum_{1\leq t<s\leq N}\sqrt{\sum_{i=1}^{n}\mathrm{K}_{\rho,\gamma}^{\alpha,\beta}(E_{\pi_{t}(i)}^{t}
+E_{\pi_{s}(i)}^{s})}\right]^{2}\right.
\nonumber\\
&\left.+L\sum_{1\leq t<s\leq N}\sum_{i=1}^{n}\mathrm{K}_{\rho,\gamma}^{\alpha,\beta}(E_{\pi_{t}(i)}^{t}+E_{\pi_{s}(i)}^{s}) +
M\sum_{1\leq t<s\leq N}\sum_{i=1}^{n}\mathrm{K}_{\rho,\gamma}^{\alpha,\beta}(E_{\pi_{t}(i)}^{t}-E_{\pi_{s}(i)}^{s}) \right\},
\end{align}
$\alpha,\beta\geq 0$, $\alpha+\beta\leq 1$, $0\leq \gamma \leq 1$, $\pi_{t},\pi_{s}\in S_n$ are arbitrary $n$-element permutations, and the parameters $L$, $M$ in $\overline{LB}1$, $\overline{LB}2$ and $\overline{LB}3$ satisfy $M\geq L>0$, $L\geq M>0$ and $L>M>0$, respectively.

According to the Appendix C in \cite{HLTG}, it is not hard to prove that our lower bound $\mathop{\mathrm{max}}\{\overline{LB}1,\overline{LB}2,\overline{LB}3\}$ is tighter than the lower bound $LB=\mathop{\mathrm{max}}\{LB1,LB2,LB3\}$ given in \cite{XWF2}.

Motivated by the results given in Appendix D of \cite{HLTG}, we have an optimal lower bound,
\begin{align}\label{eq38}
\sum_{t=1}^{N}\mathrm{K}_{\rho,\gamma}^{\alpha,\beta}(\Phi_t)\geq \mathop{\mathrm{max}}\{\overline{LB}1,\overline{LB}2,\overline{LB}_3\},
\end{align}
where $\overline{LB}_3$ is given in Theorem 2. For convenience, we denote by $\overline{LB}$ the right hand sides of (\ref{eq38}), $\overline{LB}=\mathop{\mathrm{max}}\{\overline{LB}1,
\overline{LB}2,\overline{LB}_3\}$.

To illustrate our results, we consider the MWWYD skew information as a special case, and take $M=2,L=1$ for $\overline{LB}1$, and $M=1,L=2$ for $\overline{LB}2$ and $\overline{LB}_3$.

{\bf Example 2} Given a qubit state
$\rho=\frac{1}{2}(\mathbf{1}+\mathbf{r}\cdot\bm{\sigma})$, where $\mathbf{1}$ is the $2\times2$ identity matrix, $\mathbf{r}=(\frac{\sqrt{3}}{2}\cos\theta,\frac{\sqrt{3}}{2}\sin\theta,0)$, $\bm{\sigma}=(\sigma_1,\sigma_2,\sigma_3)$
with $\sigma_j$ $(j=1,2,3)$ the Pauli matrices, and
$\mathbf{r}\cdot\bm{\sigma}=\sum^3_{j=1}r_j\sigma_j$. We consider the following three quantum channels: \\
(i) the amplitude damping channel $\Phi_{AD}$,
\begin{align*}
\Phi_{AD}(\rho)=\sum_{i=1}^2A_i\rho A_i^\dag, \quad
A_1=|0\rangle\langle0|+\sqrt{1-q}|1\rangle\langle1|, \quad A_2=\sqrt{q}|1\rangle\langle1|;
\end{align*}
(ii) the phase damping channel $\Phi_{PD}$,
\begin{align*}
\Phi_{PD}(\rho)=\sum_{i=1}^2B_i\rho B_i^\dag,\quad  B_1=|0\rangle\langle0|+\sqrt{1-q}|1\rangle\langle1|, \quad B_2=\sqrt{q}|0\rangle\langle1|;
\end{align*}
(iii) the bit flip channel $\Phi_{BF}$,
\begin{align*}
\Phi_{BF}(\rho)=\sum_{i=1}^2C_i\rho C_i^\dag,\quad  C_1=\sqrt{q}|0\rangle\langle0|+\sqrt{q}|1\rangle\langle1|, \quad C_2=\sqrt{1-q}(|0\rangle\langle1|+|1\rangle\langle0|)
\end{align*}
with $0\leq q<1$.
We compare the lower bounds $\overline{LB}$, $LB$, $LB_3$, $LB_2$, $LB_1$ with the sum $=\mathrm{K}_{\rho}^{\frac{1}{4}}(\Phi_{AD})
+\mathrm{K}_{\rho}^{\frac{1}{4}}(\Phi_{PD})+\mathrm{K}_{\rho}^{\frac{1}{4}}(\Phi_{BF})$ for $\alpha=\frac{1}{4}$, $q=0.3$. It is shown that our new lower bound $\overline{LB}$ is tighter than $LB$, $LB_3$, $LB_2$ and $LB_1$, see Figure $2$.
\begin{figure}[H]\centering
\subfigure[]
{\begin{minipage}[Xu-Cong-uncertainty2a]{0.49\linewidth}
\includegraphics[width=0.95\textwidth]{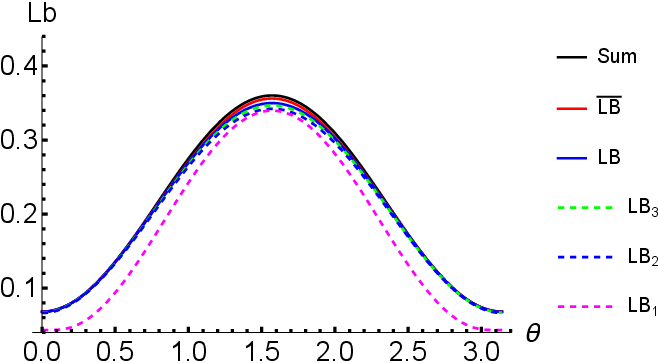}
\end{minipage}}
\subfigure[]
{\begin{minipage}[Xu-Cong-uncertainty2b]{0.49\linewidth}
\includegraphics[width=0.95\textwidth]{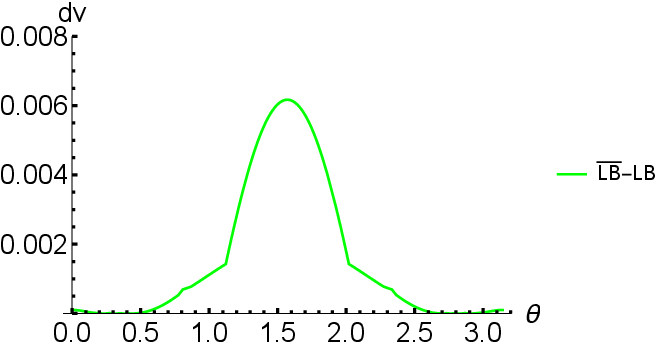}
\end{minipage}}
\subfigure[]
{\begin{minipage}[Xu-Cong-uncertainty2c]{0.49\linewidth}
\includegraphics[width=0.95\textwidth]{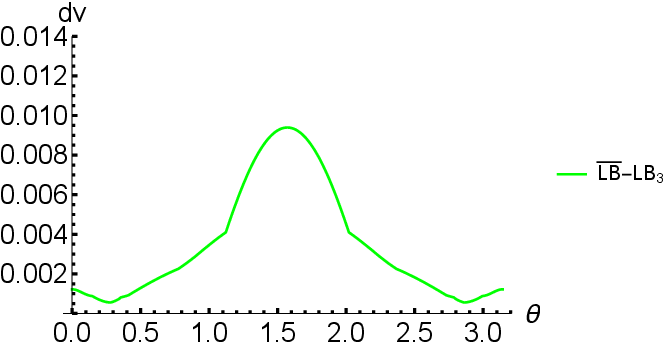}
\end{minipage}}
\subfigure[]
{\begin{minipage}[Xu-Cong-uncertainty2d]{0.49\linewidth}
\includegraphics[width=0.95\textwidth]{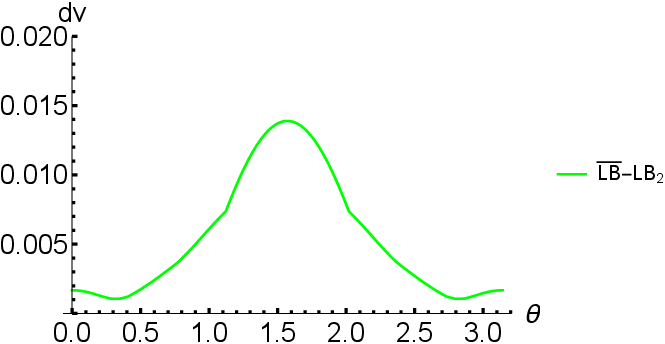}
\end{minipage}}
\caption{{(a) The solid black, solid red and solid blue curves represent the sum
$=\mathrm{K}_{\rho}^{\frac{1}{4}}(\Phi_{AD})
+\mathrm{K}_{\rho}^{\frac{1}{4}}(\Phi_{PD})+\mathrm{K}_{\rho}^{\frac{1}{4}}(\Phi_{BF})$
 , the lower bounds (Lb) $\overline{LB}$ and $LB$, respectively. The dashed green, dashed blue and dashed magenta curves represent the lower bounds $LB_3$, $LB_2$ and $LB_1$, respectively. (b-d) The solid green curve denotes the difference value (dv) between $\overline{LB}$ and $LB$, $LB_3$ and $LB_2$, respectively. \label{fig:Fig2}}}
\end{figure}

%

\medskip
\noindent {\bf 4. Sum uncertainty relations for finite unitary channels}
\\ \hspace*{\fill}\\
In this section, we explore tighter uncertainty relations of arbitrary
finite unitary channels. In \cite{XWF2}, the authors introduced the ($\alpha,\beta,\gamma$) MWWYD skew information of a unitary operator $U$,
\begin{align}\label{39}
\mathrm{K}_{\rho,\gamma}^{\alpha,\beta}(U) \notag
=&-\frac{1}{2}\mathrm{Tr}([(1-\gamma)\rho^{\alpha}
+\gamma\rho^{\beta},U^{\dag}][(1-\gamma)\rho^{\alpha}
+\gamma\rho^{\beta},U]\rho^{1-\alpha-\beta})\\
=&\frac{1}{2}\left\|\left[(1-\gamma)\rho^\alpha+\gamma\rho^\beta,
U\right]\rho^\frac{1-\alpha-\beta}{2}\right\|^{2},~~~\alpha,\beta\geq 0,~\alpha+\beta\leq 1,~0\leq\gamma \leq 1.
\end{align}

Associated with a unitary channel $\Phi_U(\rho)=U\rho U^{\dag}$,
we denote $\mathrm{K}_{\rho,\gamma}^{\alpha,\beta}(U)$ the related quantity
of the unitary channel $\Phi_U$. By employing the inequalities (\ref{eq16}), (\ref{eq17}), (\ref{eq18}), $\|\left(\sum_{t=1}^{N}u_t\right)\|^2=2\mathrm{K}_{\rho,\gamma}^{\alpha,\beta}
\left(\sum_{t=1}^{N}U_t\right)$, $\|u_t\|^2=2\mathrm{K}_{\rho,\gamma}^{\alpha,\beta}(U_t)$ and
$\|u_t\pm u_s\|^2=2
\mathrm{K}_{\rho,\gamma}^{\alpha,\beta}(U_t\pm U_s)$, we have the following theorem.

{\bf Theorem 4} Let $U_{1},\cdots,U_N$ be $N$ unitary operators, we have
\begin{align}\label{eq40}
\sum_{t=1}^{N}\mathrm{K}_{\rho,\gamma}^{\alpha,\beta}(U_t)\geq \mathop{\mathrm{max}}\{\overline{Lb}1,\overline{Lb}2,\overline{Lb}3\},
\end{align}
where
\begin{align}\label{eq41}
\overline{Lb}1
&=\frac{1}{MN+(N-2)L}\left\{\frac{2L}{N(N-1)}\left[\sum_{1\leq t<s\leq N}\sqrt{\mathrm{K}_{\rho,\gamma}^{\alpha,\beta}(U_t
+U_s)}\right]^{2}\right.
\nonumber\\
&\left.+M\sum_{1\leq t<s\leq N}\mathrm{K}_{\rho,\gamma}^{\alpha,\beta}(U_t-U_s) +
(M-L)\mathrm{K}_{\rho,\gamma}^{\alpha,\beta}
\left(\sum_{t=1}^{N}U_t\right)\right\}
\end{align}
with $M\geq L>0$,
\begin{align}\label{eq42}
\overline{Lb}2
&=\frac{1}{MN+(N-2)L}\left\{\frac{2M}{N(N-1)}\left[\sum_{1\leq t<s\leq N}\sqrt{\mathrm{K}_{\rho,\gamma}^{\alpha,\beta}(U_t
-U_s)}\right]^{2}\right.
\nonumber\\
&\left.+L\sum_{1\leq t<s\leq N}\mathrm{K}_{\rho,\gamma}^{\alpha,\beta}(U_t+U_s) +
(M-L)\mathrm{K}_{\rho,\gamma}^{\alpha,\beta}
\left(\sum_{t=1}^{N}U_t\right)\right\}
\end{align}
with $L\geq M>0$,
\begin{align}\label{eq43}
\overline{Lb}3
&=\frac{1}{MN+(N-2)L}\left\{\frac{M-L}{(N-1)^2}\left[\sum_{1\leq t<s\leq N}\sqrt{\mathrm{K}_{\rho,\gamma}^{\alpha,\beta}(U_t
+U_s)}\right]^{2}\right.
\nonumber\\
&\left.+L\sum_{1\leq t<s\leq N}\mathrm{K}_{\rho,\gamma}^{\alpha,\beta}(U_t+U_s) +
M\sum_{1\leq t<s\leq N}\mathrm{K}_{\rho,\gamma}^{\alpha,\beta}(U_t-U_s) \right\}
\end{align}
with $L>M>0$, $\alpha,\beta\geq 0$, $\alpha+\beta\leq 1$ and $0\leq \gamma \leq 1$.

For convenience, we denote by $\widetilde{Lb}$ the right hand sides of (\ref{eq40}), $\widetilde{Lb}=\mathop{\mathrm{max}}\{\overline{Lb}1,\overline{Lb}2,\overline{Lb}3\}$ and compare the lower bound with $Lb=\mathop{\mathrm{max}}\{Lb1,Lb2,Lb3\}$ in \cite{XWF2}. In the example below we consider the MWWYD skew information with $M=2,L=1$ for $\overline{Lb}1$, and $M=1,L=2$ for $\overline{Lb}2$ and $\overline{Lb}3$.

{\bf Example 3} Given a qubit state $\rho=\frac{1}{2}(\mathbf{1}+\mathbf{r}\cdot\bm{\sigma})$ with
$\mathbf{r}=(\frac{\sqrt{2}}{2}\cos\theta,\frac{\sqrt{2}}{2}\sin\theta,0)$.
Consider the following three unitary operators,
$$
U_1=e^{\frac{i\pi\sigma_1}{8}}=\left(\begin{matrix}
\cos\frac{\pi}{8}\ i\sin\frac{\pi}{8}\\
i\sin\frac{\pi}{8}\ \cos\frac{\pi}{8}
\end{matrix}
\right),
U_2=e^{\frac{i\pi\sigma_2}{8}}=\left(\begin{matrix}
\cos\frac{\pi}{8}\ \sin\frac{\pi}{8}\\
-\sin\frac{\pi}{8}\ \cos\frac{\pi}{8}
\end{matrix}
\right),
U_3=e^{\frac{i\pi\sigma_3}{8}}=\left(\begin{matrix}
e^{i\frac{\pi}{8}} \quad 0\\
\ 0\  -e^{i\frac{\pi}{8}}
\end{matrix}
\right),
$$
which correspond to the Bloch sphere rotations of $\frac{\pi}{4}$ around the $x$ axis, the $y$ axis and $z$ axis, respectively. We compare the lower bounds $\widetilde{Lb}$, $Lb$ with the sum $=\mathrm{K}_{\rho}^{\alpha}(\Phi_{AD})
+\mathrm{K}_{\rho}^{\alpha}(\Phi_{PD})+\mathrm{K}_{\rho}^{\alpha}(\Phi_{BF})$ via MWWYD skew information for $\alpha=\frac{1}{3}$ and $\alpha=\frac{1}{5}$, respectively. Our new lower bound $\widetilde{Lb}$ is tighter than $Lb$, see Figure $3$.
\begin{figure}[H]\centering
\subfigure[]
{\begin{minipage}[Xu-Cong-uncertainty3a]{0.49\linewidth}
\includegraphics[width=0.95\textwidth]{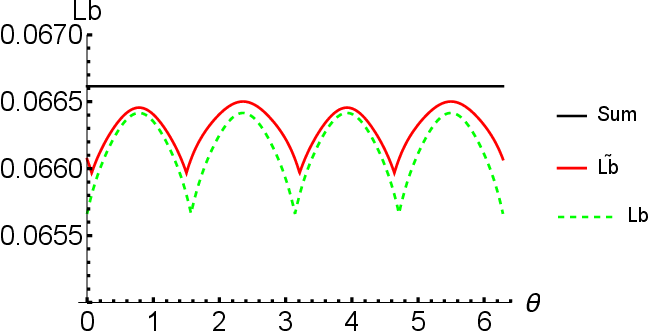}
\end{minipage}}
\subfigure[]
{\begin{minipage}[Xu-Cong-uncertainty3b]{0.49\linewidth}
\includegraphics[width=0.95\textwidth]{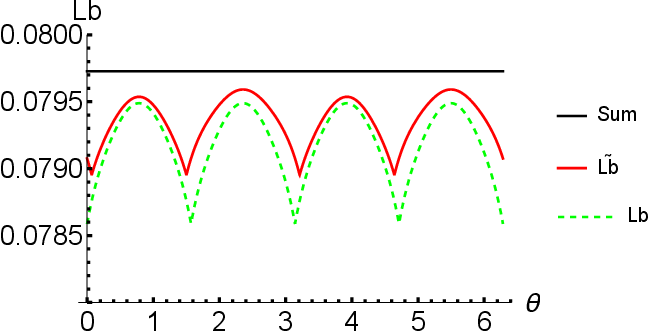}
\end{minipage}}
\caption{{The solid black, red and green dashed curves represent the sum $=\mathrm{K}_{\rho}^{\alpha}(U_1)+\mathrm{K}_{\rho}^{\alpha}(U_2)+\mathrm{K}_{\rho}^{\alpha}(U_3)$, $\widetilde{Lb}$ and $Lb$, respectively.
 (a)$\alpha=\frac{1}{3}$; (b) $\alpha=\frac{1}{5}$.  \label{fig:Fig3}}}
\end{figure}

\medskip
\noindent {\bf 5. Conclusions}\\
We have presented tighter uncertainty relations via $(\alpha,\beta,\gamma)$WWYD skew information for multiple observables, quantum channels and unitary channels. By explicit examples, we have shown that our uncertainty inequalities are tighter than the existing results given in \cite{XWF1,XWF2}. Besides, our results are also valid for the WY, WWYD and ($\alpha,\gamma$) WWYD skew information as the special cases. Uncertainty relations give rise to fundamental limitations on quantum physical quantities, and our results may shed new lights on understanding uncertainty relations and their applications in quantum information processing such as the communication security.

\vskip0.1in

\noindent

\subsubsection*{Acknowledgements}
This work was supported by National Natural Science Foundation of
China (Grant Nos. 12161056, 12075159, 12171044); Jiangxi Provincial Natural Science Foundation (Grant No. 20232ACB211003); Beijing
Natural Science Foundation (Grant No. Z190005); the Academician
Innovation Platform of Hainan Province.

\subsubsection*{Conflict of interest}
\small {The authors declare that they have no conflict of interest.}



\end{document}